\documentstyle[aps,pra,epsfig,twocolumn]{revtex}

\def\be{\begin{equation}}
\def\ee{\end{equation}}
\def\bea{\begin{eqnarray}}
\def\eea{\end{eqnarray}}
\def\bma{\begin{mathletters}}
\def\ema{\end{mathletters}}
\newcommand{\one}{\mbox{$1 \hspace{-1.0mm}  {\bf l}$}}
\newcommand{\eins}{\mbox{$1 \hspace{-1.0mm}  {\bf l}$}}
\def\C{\hbox{$\mit I$\kern-.7em$\mit C$}}

\newcommand{\ket}[1]{ | \, #1  \rangle}

\begin{document}
\draft

\title{Entangling operations and their implementation using a small amount of entanglement}

\author{J. I. Cirac$^1$, W. D\"ur$^1$, B. Kraus$^1$, and M. Lewenstein$^2$}

\address
{$^1$Institut f\"ur Theoretische Physik, Universit\"at Innsbruck,A-6020 Innsbruck, Austria\\
$^2$Institut f\"ur Theoretische Physik, Universit\"at Hannover, Hannover, Germany}

\date{\today}

\maketitle

\begin{abstract}
We study when a physical operation can produce entanglement between
two systems initially disentangled. The formalism we develop allows
to show that one can perform certain non--local operations with unit
probability by performing local measurement on states that are weakly
entangled.
\end{abstract}


\narrowtext

Much of the theoretical effort in Quantum Information Theory has been focused
so far in characterizing and quantifying the entangement properties of
multiparticles states. The reason for that lies, in part, in the fact
that those states offer interesting applications in the fields of
computation and communication. In practice, these states are created
by some physical action (or operation) involving the interaction between several systems.
This suggests that the analysis of these operations with regard to the possibility
of creating entanglement may play an important role in Quantum Information
Theory. The first steps in this direction have been recently reported
\cite{Za00,Du00e}. There, given a Hamiltonian describing the interactions
of two systems, it has been analyzed how to produce entanglement in
an optimal way.

In this letter we investigate which physical operations are capable of
producing entanglement. This goal is partly motivated by the recent
spectacular experimental progress in the field, where several physical
set--ups have been recognized to produce entangled states \cite{Fo00}.
Thus, some of the questions we analyze in this paper can  be stated as
follows: given a machine acting on two systems, can it create
entanglement? If so, what kind of entanglement? The basic mathematical
tool to answer these questions is
the isomorphism introduced by Jamiolkowski \cite{Ja75}. We will extend such an isomorphism
to relate physical operations [equivalently, completely positive maps
(CPM) ${\cal E}$] on
two systems and unnormalized states (positive operators $E$) acting
on two other systems. This allows us to reduce the problem of the
characterization of physical operations to the one of physical states.

The relation between physical operations and states has a well defined
physical meaning. In fact, from the isomorphism it follows naturally
that given a physical operation ${\cal E}$ acting on two separated
systems A and B initially disentangled (but entangled locally to
some other ancilla systems) we can always obtain the corresponding
state $E$ as an outcome. What is even more surprising is that, starting from the state
$E$ we can always perform some local measurements such that for
certain outcomes the state of systems A and B changes exactly
as if we had applied the corresponding operation ${\cal E}$.

This last
property will allow us to answer an intriguing question raised in
the context of Quantum Information Theory. Let us assume that
we have two qubits A and B at different locations and we want to apply
some non--local operation. This situation raises, for example, in
the context of distributed quantum computation \cite{Ci98}, where non--local
operations between different quantum computers are required.
So far, it is known that one can use maximally entangled states, local operations and classical communication (LOCC) to perform
that task as follows: we can teleport the state of A to the location of
B, perform the operation locally, and then teleport the corresponding
state back to A. In this process one has to consume two maximally entangled
states (i.e. two ebits) apart from transmitting two classical bits in
each direction \cite{Ch00}. However, it
is known that for some kind of operations (like the controlled--NOT
gate) one can economize the resources, such that only one ebit is
consumed \cite{Go98}. In fact, all the operations that have been
studied so far \cite{Ch00,Go98,Ei00,Co00}
require {\em an integer number of ebits}. We will show here that
many operations require {\em a non--integer number of ebits}. In
particular, if the operation can only entangle qubits weakly, the required number
is much smaller than one. This automatically implies that many
tasks in distributed quantum computation can be performed with a
much smaller entanglement than the one required so far.

Let us consider two systems A and B at different locations,
whose states are represented
by vectors in the Hilbert space ${\cal H}_{A,B}$, respectively, both of dimension
$d$. Any physical
action on those systems is represented mathematically by a completely
positive linear map ${\cal E}$ mapping the density operator $\rho$ of
those systems onto another positive operator ${\cal E}(\rho)$. The map can be written as
\be
{\cal E}(\rho) = \sum_k O_k \rho O_k^{\dagger},
\ee
where $O_k$ are operators acting on ${\cal H}_A\otimes {\cal H}_B$. For the sake
of generality, we have not imposed that the map preserves the trace
of $\rho$, since we may be interested in physical actions that
occur with certain probability \cite{Sc96}.

Our first goal is to determine when a given CPM is able to produce
entangled states. Thus, we first recall the definition of separable
operators. We say that a density operator
$\rho$ is separable with respect to systems A and B if it can be written as
\cite{We89}
\be
\label{rhos}
\rho = \sum_{i=1}^n |a_i\rangle_A\langle a_i| \otimes |b_i\rangle_B\langle b_i|,
\ee
for some integer $n$, and
$\ket{a_i}_A\in {\cal H}_A$ and $\ket{b_i}_B\in {\cal H}_B$. Otherwise we say that
it is non--separable (or entangled).
Separable positive operators describe
states that can be prepared using local operations and classical
communication out of product states, i.e. are useless for quantum information
tasks that require entanglement. During the last years, much theoretical
effort has been devoted to study the separability properties of operators
\cite{separability}. In particular, a necessary condition for separability of a given
positive operator $\rho$ is that $\rho^{T_A}\ge 0$ \cite{Pe96,Ho96}, where $T_A$ denotes
transposition in ${\cal H}_A$ in a given orthonormal basis $S_A=\{|k\rangle\}_{k=1}^d$.
This condition turns out to be sufficient as
well when the sum of the dimensions of ${\cal H}_{A,B}$ does not exceed five (for
example, for two qubits). In higher dimension there are examples of
entangled states represented by non--separable
operators whose partial transpose is positive \cite{Ho97}. For methods to study
separability of operators which have positive partial transposition
we refer the reader to \cite{separability}.

We can similarly define separable CPM; that is, ${\cal E}$ is separable
\cite{Ra98}
if its action can be expressed in the form
\be
\label{Es}
{\cal E}(\rho) = \sum_{i=1}^n (A_i\otimes B_i) \rho
(A_i\otimes B_i)^\dagger,
\ee
for some integer $n$ and where $A_i$ and $B_i$ are operators acting
on ${\cal H}_{A,B}$, respectively. Otherwise, we say that it is non--separable.
Up to a proportionality constant, separable
maps are those that can be implemented using local
operations and classical communication only \cite{note}, i.e. useless for several
tasks in quantum information.

From the defintions (\ref{rhos}) and (\ref{Es}) it follows that
if ${\cal E}$ and $\rho$ are
separable, then ${\cal E}(\rho)$ is also separable. This reflects the fact
that by local actions one cannot create entanglement.

Let us consider a CPM ${\cal E}$ acting on systems A$_1$ and B$_1$.
We define the operator $E_{A_1A_2,B_1B_2}$ acting on
${\cal H}_A\otimes {\cal H}_B$ [where now ${\cal H}_A={\cal H}_{A_1}\otimes {\cal H}_{A_2}$ and ${\cal H}_B={\cal H}_{B_1}\otimes {\cal H}_{B_2}$,
and ${\rm dim}({\cal H}_{A_i})={\rm dim}({\cal H}_{B_i})=d$]
as follows:
\be
\label{EAB}
E_{A_1A_2,B_1B_2} = {\cal E}(P_{A_1A_2}\otimes P_{B_1B_2}).
\ee
Here, $P_{A_1A_2}=|\Phi\rangle_{A_1A_2}\langle \Phi|$ with
\be
\label{MES}
|\Phi\rangle_{A_1A_2} = \frac{1}{\sqrt{d}}\sum_{i=1}^d |i\rangle_{A_1}\otimes
|i\rangle_{A_2},
\ee
and $S=\{|i\rangle\}_{i=1}^d$ an orthonormal basis. In the definition
(\ref{EAB}) the map ${\cal E}$ is understood to act as the identity
on the operators acting on ${\cal H}_{A_2}$ and ${\cal H}_{B_2}$. The operator $E$
has a clear interpretation since it is proportional to the density
operator resulting from the operation ${\cal E}$ on systems
A$_1$ and B$_1$ when both of them are prepared in a maximally entangled state with
two ancillary systems, respectively.

On the other hand, we have
\be
\label{ECPM}
{\cal E}(\rho_{A_1B_1})= d^4{\rm tr}_{A_2A_3B_2B_3}(E_{A_1A_2,B_1B_2}
\rho_{A_3B_3} P_{A_2A_3}P_{B_2B_3}).
\ee
This can be proved as follows. First, we can write
\be
d^2{\rm tr}_{A_3B_3} (\rho_{A_3B_3} P_{A_2A_3}P_{B_2B_3})=\rho_{A_2B_2}^T,
\ee
where $T$ means transpose in the basis $S_{A_2}\otimes S_{B_2}$.
Now, using (\ref{EAB}) one can readily show that
\be
{\cal E}(\rho_{A_1B_1})= d^2{\rm tr}_{A_2B_2}(E_{A_1A_2,B_1B_2} \rho_{A_2B_2}^T).
\ee
Equation (\ref{ECPM}) has a very simple interpretation. It reflects the
fact that if we
have the state $E_{A_1A_2,B_1B_2}$ at our disposal, we can always produce
the map ${\cal E}$ on any state of systems A$_3$ and B$_3$ by performing a joint measurement
locally such that both systems A$_2$A$_3$ and  B$_2$B$_3$ are projected onto the
maximally entangled state ({\ref{MES}). Of course, this will happen with
certain probability. Below we will show how to implement CPM with
unit probability using this method.

The relations (\ref{EAB}) and (\ref{ECPM}) induce a one--to--one
correspondence between CPM acting on tensor product spaces and
positive operators. In fact, this correspondece can be viewed as an extension of the
isomorphism introduced by Jamiolkowski \cite{Ja75} to tensor product spaces.
Using these relation it is very easy to show that:

\begin{description}

\item [(i)] ${\cal E}$ is separable iff $E_{A_1A_2,B_1B_2}$ is separable
with respect to the systems (A$_1$A$_2$) and (B$_1$B$_2$). Thus, we
can study the separability of
CPM by studying the problem of separability of positive operators. This immediately
implies that we can use all the results derived for the latter problem \cite{separability}.

\item [(ii)] ${\cal E}$ can create entanglement iff $E_{A_1A_2,B_1B_2}$ is non--separable
with respect to the systems (A$_1$A$_2$) and (B$_1$B$_2$). In particular, we can always
obtain a state whose density operator is proportional to $E_{A_1A_2,B_1B_1}$ out
of separable states by
entangling our systems locally with ancillas.

\item [(iii)] Let us assume that
$E_{A_1A_2,B_1B_2}^{T_{A_1A_2}}\ge 0$, where $T_{A_1A_2}$ denotes transposition
with respect to A$_1$ and A$_2$ in the basis $S_A$. Then, if $\rho^{T_{A_1}}\ge 0$ we
have that ${\cal E}(\rho_{A_1B_1})^{T_{A_1}}\ge 0$.
If additionally $E_{A_1A_2,B_1B_2}$ is entangled (i.e. bound entangled), then
we can always produce bound entangled states out of non--entangled states
by using the map ${\cal E}$. We
just have to entangle the systems locally with ancillas.

\item [(iv)] If ${\cal E}$ corresponds to a unitary action, the
corresponding operator has rank one, i.e. it can be written as
$E= |\Psi\rangle\langle \Psi|$, where $|\Psi\rangle\in {\cal H}_{A_1}\otimes
{\cal H}_{A_2}\otimes {\cal H}_{B_1}\otimes {\cal H}_{B_2}$ is a normalized state.

\end{description}

Let us consider some simple examples with qubits ($d=2$).
First, let
us assume that $E_{A_1A_2,B_1B_2}$ is an entangled state with
positive partial transposition. According to
(i) the corresponding completely positive map ${\cal
E}$ is nonseparable and according (iii) $[{\cal
E}(\rho)]^{T_{A_1}}\ge 0$ for all $\rho$ separable. But in this
case, positive partial transposition is equivalent to separability
\cite{Pe96,Ho96}, and therefore ${\cal E}(\rho)$ is separable for all $\rho$
separable. However, if we allow for input states that are locally
entangled with ancillas, the final state will be (bound) entangled
according to (ii).

On the other hand, let us
consider a family of phase gates of the form
\be
\label{UN}
U(\alpha_N) \equiv e^{-i\alpha_N \sigma_x^{A_1}\otimes\sigma_x^{B_1}},
\mbox{    }\alpha_N\equiv \pi/2^N\label{Ualpha},
\ee
where the $\sigma$'s are Pauli operators.
These gates are of the same kind as the ones used in the discrete
Fourier transform \cite{So95}.
The corresponding operator $E_{A_1A_2,B_1B_2}= |\psi_{\alpha_N}\rangle
\langle \psi_{\alpha_N}|$, where
\bea
|\psi_{\alpha_N}\rangle &=& \cos(\alpha_N)|\Phi^+\rangle_{A_1A_2}
|\Phi^+\rangle_{B_1B_2}\nonumber\\
&&- i\sin(\alpha_N)|\Psi^+\rangle_{A_1A_2}
|\Psi^+\rangle_{B_1B_2},
\label{psialpha}
\eea
and $|\Phi^+\rangle$ and $|\Psi^+\rangle$ are Bell states.

In the following, we will use the formalism introduced above to
study how to perform non--local operations using a small amount
of entanglement.  Let us consider a basis of maximally entangled
states of systems A$_1$A$_2$ (and B$_1$B$_2$) as
$|\Phi_i\rangle=\eins\otimes U_i|\Phi\rangle$,
where $U_i$ are a unitary operators and $|\Phi\rangle$ is defined
in (\ref{MES}).
If we perform a measurement in that basis and obtain
$|\Phi_i\rangle_{A_1A_2}$ and $|\Phi_j\rangle_{B_1B_2}$ the state
of our systems will be ${\cal E}(U_i\otimes U_j \rho_{A_1B_1}U_i^\dagger
\otimes U_j^{\dagger}) $. Thus, we see that as a result of the measurement
we either implement the CPM, ${\cal E}$, or local operations followed by
the CPM. Now we will show how to use this effect in order to
perform non--local operations by using entangled states. We will
restrict ourselves to the case of qubits, but our results can be
easily generalized.

Let us start considering the gates $U(\alpha_N)$ (\ref{UN}).
The amount of entanglement of the corresponding state $|\psi_{\alpha_N}\rangle$
(\ref{psialpha}) is given by its
entropy of entanglement
\be
E(\psi_{\alpha_N})=-x \log_2(x)-(1-x)\log_2(1-x),\label{EOE}
\ee
where $x=\cos^2(\alpha_N)=\cos^2(\pi/2^N)$.
On the one hand, $E(\psi_{\alpha_2})=1$, i.e. according to our
discussion $U(\pi/4)$ is capable of creating 1 ebit of entanglement.
On the other hand,
$E(\psi_{\alpha_1})=0$, i.e. $U(\pi/2)=-i \sigma_x\otimes \sigma_x$ is
a {\it local} gate. For $N\geq 2$, we have that $E(\psi_{\alpha_N})$
is monotonically decreasing with $N$. Note that
for $N$ sufficiently large, we can
regard (\ref{UN}) as an infinitesimal transformation and use
the results of Ref.\ \cite{Du00e} to show that the gate can
optimally create an entanglement proportional to $\alpha_N$.
We will show that in that limit $U(\alpha_N)$ can be
implemented with unit probability by using an average amount of entanglement
also proportional to $\alpha_N$,
assisted by classical communication of approximately $2$ bits in both directions.
Thus, we provide examples of non--local gates which can be implemented using
much less than 1 ebit of entanglement, the required entanglement being proportional
to the entanglement capability of the non--local gate.

We want to perform the gate on systems A$_3$B$_3$ and obtain the output state
in systems A$_1$B$_1$. We assume that both systems A$_1$A$_2$ and B$_1$B$_2$
are in the state $|\psi_{\alpha_N}\rangle$
and we measure systems A$_2$A$_3$ and B$_2$B$_3$ in the Bell
basis $|\Psi_{i_1,i_2}\rangle = \eins\otimes \sigma_{i_1,i_2}|\Psi\rangle$,
where $\sigma_{1,1}=\eins,\sigma_{1,2}=\sigma_x,\sigma_{2,1}=
\sigma_y$, and $\sigma_{2,2}=\sigma_z$.
Note that all outcomes of the measurement are equally probable.
If the outcome for A$_2$A$_3$ is
$|\Psi_{i_1,i_2}\rangle$, we apply $\sigma_{i_1,i_2}$ to A$_1$ and proceed
analogously with B$_2$B$_3$.
One can readily see that the resulting operation on A$_1$B$_1$ after this procedure will be: (i)
$U(\alpha_N)$ if $i_1=j_1$; (ii) $U(\alpha_N)^\dagger=U(-\alpha_N)$ if $i_1\ne j_1$.
Thus, with probability 1/2 we obtain the desired gate, whereas with probability
1/2 we apply $U(-\alpha_N)$ instead, and so we fail. In order to apply the desired gate
with probability one, we proceed as follows. If we fail, we repeat the procedure
but with systems A$_1$A$_2$ and B$_1$B$_2$ prepared in the state $|\psi_{2\alpha_N}\rangle$. With
a probability 1/2 we will succeed, and otherwise we will have applied $U(-\alpha_N)^3$
to the original state. We continue in the same vein; that is, in the $k$--th
step we use systems A$_1$A$_2$ and B$_1$B$_2$ in the state
$|\psi_{2^{k-1}\alpha_N}\rangle$ so that
if we fail altogether we will have applied $U(-\alpha_N)^{2^{k}-1}$. For $k=N$
we have that $U(-\alpha_N)^{2^{N}-1}=-U(\alpha_N)$, and therefore even if
we fail we will have applied the right gate, so that the procedure ends.

The total average entanglement which is consumed during this procedure is given by
\be
\label{avent}
\bar E[U(\alpha_N)] = \sum_{k=1}^{N} \left(\frac{1}{2}\right)^{k-1}E(\psi_{\alpha_{N-k+1}})
= \alpha_N f_N,
\ee
where
\be
f_N = \frac{1}{\pi} \sum_{k=1}^{N} 2^k E(\psi_{\alpha_{k}})< f_{\infty}=5.97932.
\ee
In (\ref{avent}),
the weight factor of $p_k=(1/2)^{k-1}$ gives the probability that the
$k$--th step has to be performed. Thus, we obtain
$\bar E[U(\alpha_N)]< \alpha_N f_\infty$.
Due to the fact that in each step of this procedure one bit of classical
communication in each direction is necessary \cite{note3}, the average amount of classical
communication is given by $2-(1/2)^{N-2}$ bits.

Although the procedure described above allows only to implement gates with ``binary phases''
$\alpha_N=\pi/2^N$, any gate $U(\alpha)$ with arbitrary phase $\alpha$ can be
approximated with arbitrarily high accurancy by a sequence of gates of the form
$U(\alpha_N)$, consuming on average $\bar E \leq f_\infty \alpha$ ebits
of entanglement. Furthermore, this procedure allows to implement any
arbitrary two--qubit unitary operation $U$. We can write any such operation
as $U=e^{-iHt}=\lim_{n\to\infty} (\one - i H t/n)^n$, where $H$ is a
self--adjoint operator. We can thus
apply infinitesimal gates $U_n=(\one - i H t/n)$ sequentially using an extension
of the scheme described above. Note that after such an infinitesimal
operation we can perform local operations without consuming entanglement. This
allows us to restrict the form of the Hamiltonians to those that can be
written as
\be
\label{Htilde}
H_0 = \sum_{k=x,y,z}^3 \mu_k \sigma^A_k \otimes
\sigma^B_k \equiv \sum_{k=1}^3 H_k.
\ee
This can be viewed as follows. First, let us write $H$ in terms of
Pauli operators for systems A and B as
\be
H= \vec \alpha \cdot \vec\sigma^A + \vec \beta \cdot \vec\sigma^B
+ \vec \sigma^A \cdot \gamma \vec \sigma^B,
\ee
where $\gamma$ is a matrix, and $\vec \sigma$ is the Pauli vector.
If we apply an infinitesimal local transformation in A and B
with Hamiltonians $-\vec \alpha \cdot \vec\sigma^A$ and $-\vec \beta \cdot \vec\sigma^B$
respectively, this will be equivalent to having $H$ with $\alpha=\beta=0$.
Moreover,  prior to this operation and after the application of $U_n$
we can always perform local operations such that we obtain an evolution
given by $H_0$ (\ref{Htilde}), where the $\mu$'s are the singular
values of $H_0$. Since
the $H_k$ commute, we have that the corresponding unitary operation is given by
\be
\tilde U_n=e^{-i H_1t/n}e^{-i H_2 t/n}e^{-i H_3 t/n},
\ee
a sequence of operations of the form (\ref{Ualpha}), for which we already have
provided
a protocol. The required
amount of entanglement is therefore given by
$\bar E_{U}=f_\infty t (\mu_1+\mu_2+\mu_3)$ ebits.

Using the results of Ref. \cite{Du00e}, one can compare for small $\alpha_N$ (large $N$)
the average amount of entanglement used up to implement the gate (\ref{UN}) with the maximal
amount of entanglement which can be produced with help of a single application
of the gate \cite{note2}. One finds that for $\alpha_N
\rightarrow 0$ that the ratio $\bar E[U(\alpha_N)]/E_{\rm create}
[U(\alpha_N)]$ is given by $\approx 3.1268$, i.e. the amount of entanglement required
to perform the gate is about 3 times the amount of entanglement which can be created using this gate.
Similar results are found for a general $U=e^{-iHt}$ in the limit $t\to 0$.

As we have restricted ourselves to single applications of the unitary operation,
the protocol given here is very unlikely to be optimal in terms of the
consumed entanglement per application of the operation $U$.
One might expect that in some asymptotic limit, the average amount of
entanglement required to implement a gate $U$ equals the amount of
entanglement which can be produced using this gate. For $U(\pi/4)$ (\ref{Ualpha}),
we have that the protocol described above is optimal, as it consumes only 1 ebit
of entanglement, which equals the maximal amount of entanglement which can be
created with a single application of $U(\pi/4)$. Several other examples of
this kind ---all of them dealing with an integer number of ebits--- have been
proven to be optimal in \cite{Ch00,Ei00,Co00}.

Finally, let us mention that we have restricted ourselves here to the
implementation of non--local unitary operations. In fact, with the formalism
introduce here one can extend
the analysis to non--unitary operations and even to the implementation
of non--local measurements. All these results indicate that the entanglement
properties of a physical operation ${\cal E}$ are directly related to
the entanglement of the corresponding operator $E$.

%-------------------------------------------------------------------------

This work was
supported by the Austrian SF, the DFG,
%under the SFB ``control and measurement of coherent quantum systems'' (Project 11),
the European Community
under the TMR network ERB--FMRX--CT96--0087 and project EQUIP (contract
IST-1999-11053), the ESF, and the Institute for Quantum Information GmbH.

% -------------------------------------------------------------


\begin{references}
\bibitem{note}
Note that this does not contradict the statement given in Ref.\
\cite{Be98}, in which it is shown that there are certain trace
preserving separable completely positive maps that cannot be
implemented locally with probability one.
\bibitem{note3}
In the procedure described above, only $(N-2)$ steps have to be performed in practice,
as it happens that the required operation to be implemented in step $(N-1)$, $U(\pi/2)$, is a local
 operation which can be performed with certainty and without classical communication.
\bibitem{note2}
As shown in \cite{Du00e}, this amount is strictly larger
than the entanglement of the state associated to the operation, i.e. it is
better to start with some initially entangled state rather than a product
state in order to optimally increase the amount of entanglement. In the limit
$\alpha_N \rightarrow 0$, the maximal amount of created entanglement is given
by $E_{\rm create}[U(\alpha_N)]=1.9123 \alpha_N$.
\end{references}
\end{document}